\DocumentMetadata{}
\documentclass[sigplan,10pt]{acmart}
\usepackage[]{hyperref}

\usepackage[utf8]{inputenc}
\usepackage[russian, english]{babel}
\usepackage[normalem]{ulem}
\usepackage[dvipsnames]{xcolor}
\usepackage{xspace}
\usepackage{graphicx}
\usepackage{subcaption}
\usepackage{url}
\usepackage{xurl}
\usepackage{listings}
\usepackage{epigraph}

\lstdefinestyle{codestyle}{
    commentstyle=\color{green},
    keywordstyle=\color{blue},
    numberstyle=\tiny\color{gray},
    stringstyle=\color{purple},
    basicstyle=\fontsize{7.5}{8.5}\ttfamily,
    breakatwhitespace=false,         
    breaklines=true,                 
    captionpos=b,                    
    keepspaces=true,                 
    numbers=left,  
    xleftmargin=1em,
    numbersep=5pt,                  
    showspaces=false,                
    showstringspaces=false,
    showtabs=false,                  
    tabsize=2,
    frame=none,
    escapechar={|}
}

\definecolor{communication}{RGB}{86, 81, 126}
\definecolor{computation}{RGB}{174, 65, 50}

\newif\ifanonymous
\newif\ifsubmit    
\submitfalse       
\anonymousfalse    

\ifsubmit
\newcommand{\fixme}[1]{}
\newcommand{\ana}[1]{}
\newcommand{\tom}[1]{}
\newcommand{\andrea}[1]{}
\newcommand{\leon}[1]{}
\newcommand{\pinghe}[1]{}
\newcommand{\lazar}[1]{}
\newcommand{\todo}[1]{}

\newcommand{\cut}[1]{}
\else
\newcommand{\fixme}[1]{{\textcolor{red}{[~FIXME:~#1~]}}}
\newcommand{\ana}[1]{{\textcolor{orange}{[~ANA:~#1~]}}}
\newcommand{\tom}[1]{{\textcolor{purple}{[~TOM:~#1~]}}}

\newcommand{\andrea}[1]{{\textcolor{blue}{[~ANDREA:~#1~]}}}
\newcommand{\leon}[1]{{\textcolor{cyan}{[~LEON:~#1~]}}}
\newcommand{\pinghe}[1]{{\textcolor{pink}{[~PINGHE:~#1~]}}}
\newcommand{\lazar}[1]{{\textcolor{brown}{[~LAZAR:~#1~]}}}
\newcommand{\todo}[1]{{\textcolor{red}{[~TODO:~#1~]}}}

\newcommand{\cut}[1]{\sout{#1}}
\fi

\ifanonymous
\newcommand{\dandelion}{{Hummingbird}\xspace}
\newcommand{\hybrid}{HB-hybrid\xspace}
\else
\newcommand{\dandelion}{{Dandelion}\xspace}
\newcommand{\hybrid}{D-hybrid\xspace}
\fi

\newcommand{\dlibc}{{\texttt{dlibc}}\xspace}
\newcommand{\dlibcpp}{{\texttt{dlibc++}}\xspace}

\ifanonymous

\else

\fi

\newcommand{\fakepara}[1]{\noindent\textbf{#1}\hspace{2mm}}

\copyrightyear{2025}
\acmYear{2025}
\setcopyright{cc}
\setcctype{by}
\acmConference[SOSP '25]{ACM SIGOPS 31st Symposium on Operating Systems Principles}{October 13--16, 2025}{Seoul, Republic of Korea}
\acmBooktitle{ACM SIGOPS 31st Symposium on Operating Systems Principles (SOSP '25), October 13--16, 2025, Seoul, Republic of Korea}
\acmDOI{10.1145/3731569.3764803}
\acmISBN{979-8-4007-1870-0/2025/10}
\acmPrice{}


\title{Unlocking True Elasticity for the Cloud-Native Era with Dandelion}

\begin{document}

\settopmatter{authorsperrow=4}

\ifanonymous
\else
\author{Tom Kuchler}
\affiliation{\institution{ETH Zurich}\country{Switzerland}}

\author{Pinghe Li}
\affiliation{\institution{ETH Zurich}\country{Switzerland}}

\author{Yazhuo Zhang}
\affiliation{\institution{ETH Zurich}\country{Switzerland}}

\author{Lazar Cvetković}
\affiliation{\institution{ETH Zurich}\country{Switzerland}}

\author{Boris Goranov}
\affiliation{\institution{ETH Zurich}\country{Switzerland}}

\author{Tobias Stocker}
\affiliation{\institution{ETH Zurich}\country{Switzerland}}

\author{Leon Thomm}
\affiliation{\institution{ETH Zurich}\country{Switzerland}}

\author{Simone Kalbermatter}
\affiliation{\institution{ETH Zurich}\country{Switzerland}}

\author{Tim Notter}
\affiliation{\institution{ETH Zurich}\country{Switzerland}}

\author{Andrea Lattuada}
\affiliation{\institution{MPI-SWS}\country{Germany}}

\author{Ana Klimovic}
\affiliation{\institution{ETH Zurich}\country{Switzerland}}

\begin{abstract}

Elasticity is fundamental to cloud computing. An elastic platform can quickly allocate resources to match the demand of each workload as it arrives, rather than pre-provisioning resources to meet performance objectives. However, even serverless platforms --- which boot sandboxes in 10s to 100s of milliseconds --- are not sufficiently elastic to avoid pre-provisioning expensive resources. Today's FaaS platforms provision many extra, idle sandboxes in memory to reduce the occurrence of slow, cold starts. Initializing securely isolated sandboxes with a POSIX-like computing environment that today's cloud users expect is slow as it requires booting a guest OS and configuring networking.

Our key insight is that the rise of cloud-native application development provides an opportunity to rethink the application interface to the cloud and co-design a much more efficient, elastic computing platform under the hood.
We propose \dandelion, an elastic cloud platform with a declarative cloud-native programming model that replaces POSIX-based network interfaces with higher-level (e.g., HTTP-based) interfaces for applications to interact with remote services like cloud storage, databases, and AI inference services. \dandelion executes applications expressed as DAGs of pure compute functions and communication functions.

This enables \dandelion to securely execute compute functions in lightweight sandboxes that cold start in \textit{100s of microseconds}, since pure functions do not rely on software environments such as a guest OS.
\dandelion makes it practical to boot sandboxes on-demand per request, decreasing performance variability by two to three orders of magnitude compared to Firecracker and reducing committed memory by 96\% on average when running the Azure Functions trace.

\end{abstract}

\maketitle

\renewcommand{\shortauthors}{Kuchler et al.}

\pagenumbering{gobble}

\section{Introduction}\label{sec:intro}

\epigraph{\textit{\textbf{Elasticity} is the degree to which a system is able to adapt to workload changes by provisioning and deprovisioning resources in an autonomic manner, such that at each point in time, the available resources match the current demand as closely as possible~\cite{elasticity-definition}.}}{}

Elasticity is a key selling point of cloud computing~\cite{view-of-cloud-2010}.
The more elastic the infrastructure, the more cost-efficient cloud computing can become, since fewer resources need to be pre-provisioned before a request arrives to ensure low latency.
Infrastructure as a Service (IaaS) offers greater elasticity than purchasing on-premise servers, as users can boot virtual machines (VMs) in tens of seconds~\cite{boxer-elasticity} and rent them for as long as they need. Serverless computing~\cite{RiseofServerless19} pushes elasticity even further, as Function as a Service (FaaS) platforms boot tasks in tens to hundreds of milliseconds~\cite{firecracker,reap, catalyzer} and autoscale sandboxes based on invocation load~\cite{serverless-in-the-wild}.

However, even serverless cloud platforms are not elastic enough to avoid pre-provisioning expensive resources. To achieve low latency and high throughput, FaaS platforms keep many idle function sandboxes ready in memory~\cite{firecracker, serverless-in-the-wild}. For example, when running the Azure Functions trace~\cite{serverless-in-the-wild}, the Knative~\cite{knative} serverless platform aggressively autoscales sandboxes such that 97\% of requests execute in sandboxes whose memory was provisioned before the invocation arrived (warm starts).  
Although this improves median latency, it causes the platform to commit 16$\times$ more memory on average (red line in Figure~\ref{fig:mem-overhead}) than the memory required for running active requests (blue line). Provisioning so many extra sandboxes in memory is expensive, since DRAM dominates server costs~\cite{tmo-dram-cost, pond-dram-cost}. 

\begin{figure}
	\centering
	\includegraphics[trim={0 1.0cm 0 0.5cm},width=0.5\textwidth]{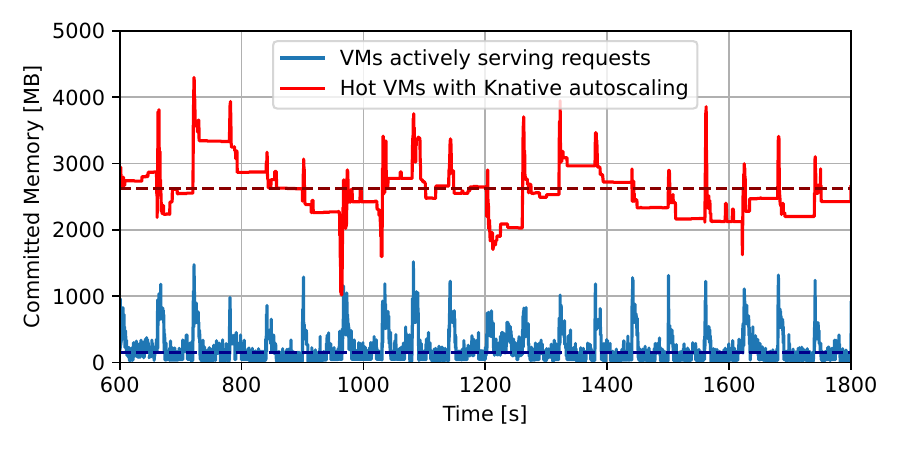}
	\caption{Azure trace execution with 100 functions, comparing committed memory for all VMs on the node with Knative autoscaling vs. for the VMs actively serving requests. Keeping VMs hot reduces cold starts, but consumes extra DRAM, which is expensive. Dotted lines show the average.}
	\label{fig:mem-overhead}
\end{figure}

The current approach of capping costs by pre-provisioning sandboxes for many, but not all functions leads to the additional problem of high performance variability. The small percentage of requests that are cold starts incur noticeable extra latency to boot a sandbox on the critical path. Figure~\ref{fig:hotStartServer} shows how sensitive tail latency is to the percentage of hot requests (note the log scale!), despite sandbox creation optimizations like snapshots~\cite{firecracker:snapshot_restoration}. The high tail latency due to cold starts particularly impacts user-facing tasks like web serving and interactive data processing. 

We show that a key barrier to achieving true elasticity in the cloud is executing each function in a sandbox derived from a container or VM that \textit{exposes a POSIX-like interface to the function}. 
A minimal KVM~\cite{kvm} instance can boot in less than 140 thousand cycles~\cite{virtines} (i.e. $\sim52\mu s$ at 2.69GHz). However, booting a hello-world function in a Firecracker MicroVM (which is based on KVM) takes over 10\textit{ms} even with snapshots.
Much of this time is spent on loading the guest OS snapshot and configuring the MicroVM's network setup (>8\textit{ms} is spent on demand paging the snapshot and re-establishing the guest-host connection), which are needed to provide a POSIX-like interface to the user function. 

Although prioritizing POSIX compatibility was essential in the early days of cloud computing to ease migration from on-premise deployments, cloud-native application development is increasingly common today. Developers build applications for cloud environments, leveraging numerous cloud services for storage, analytics, AI inference, and more, which are all exposed over REST APIs~\cite{aws-rest-api}. Cloud-native programming is evolving~\cite{serverless-end-linux, new-directions-cloud, rethinking-serverless, compute-centric-networking, func_as_func, mosaic, pulumi, yu2023following, cloudburst, boxer-elasticity, serverless-uphill, lambdata, dataflower}, with systems like SigmaOS~\cite{sigmaos} demonstrating that well-tailored interfaces for cloud-native applications can eliminate the need for traditional POSIX-like interfaces. Even today's FaaS platforms, such as AWS Lambda, are not fully POSIX compliant, as functions cannot accept incoming connections~\cite{wawrzoniak2021boxer}. 
Hence, now is the right time to rethink cloud abstractions that can enable the underlying infrastructure to become more efficient and truly elastic. 

\begin{figure}
	\centering
	\includegraphics[trim={0 1.25cm 0 0},width=\columnwidth]{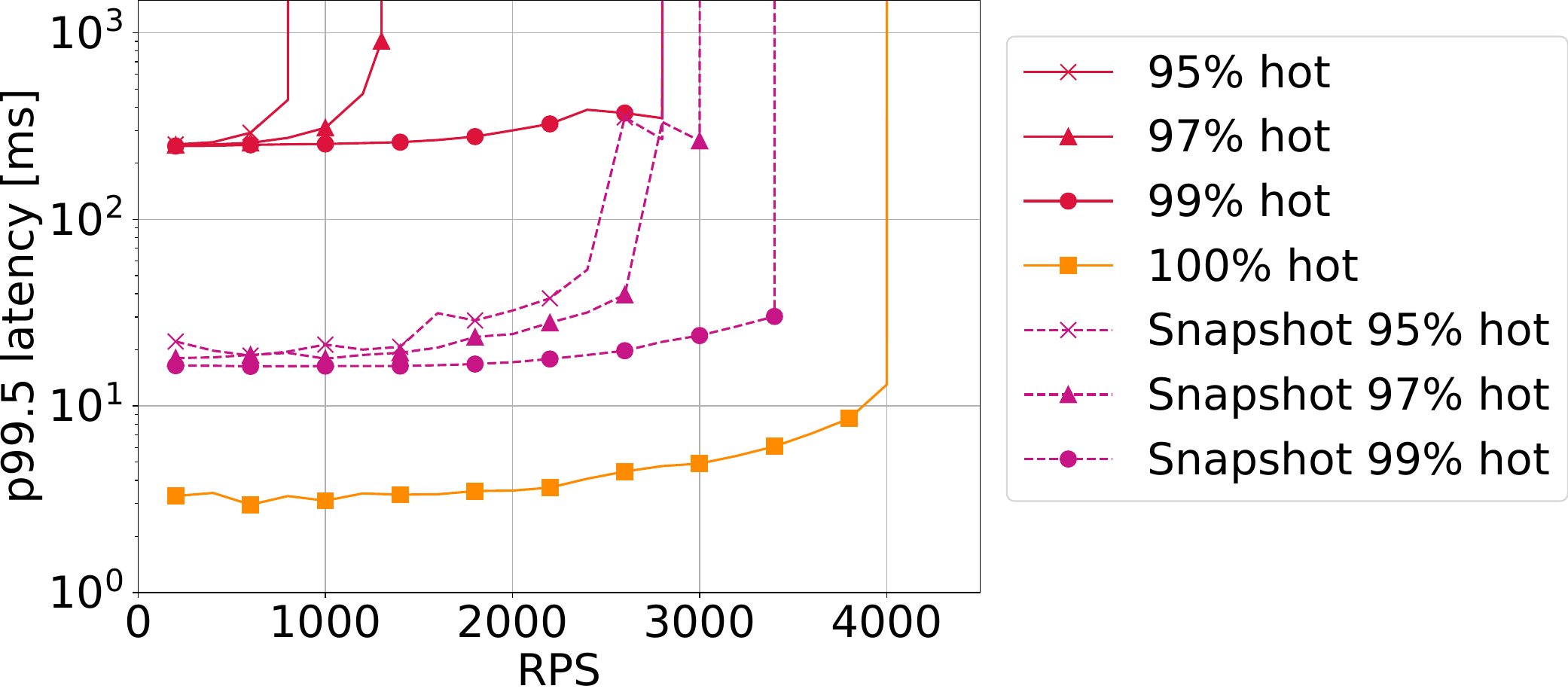}
	\caption{128x128 int64 matmul running in Firecracker MicroVMs. The \% of cold requests greatly impacts performance.}
	\label{fig:hotStartServer}
\end{figure}

We propose \dandelion, an elastic cloud platform with a declarative cloud-native programming model that replaces the POSIX-based network interface with a higher-level (e.g., HTTP-based) interface for applications to interact with remote services like cloud storage, databases, and AI inference services. \dandelion executes applications expressed as direct acyclic graphs (DAGs) of \textit{pure compute functions} and \textit{communication functions}. Pure compute functions are bodies of (untrusted) user code, which take a declared list of inputs and produce a declared list of outputs. Communication functions are implemented by the \dandelion platform and can be invoked (but not modified) in user application DAGs. \dandelion currently implements a communication function for HTTP communication, which enables REST API interactions. We plan to add more communication functions to support additional protocols.

Decomposing a cloud application into pure compute and communication functions enables redesigning the function execution system for high elasticity.
First, \dandelion can securely execute pure compute functions in lightweight sandboxes that are fast to cold start, since the sandboxes do not need a guest OS setup for networking, threading, etc. or an isolated virtual network device on the host.
Second, \dandelion can dedicate a CPU core for each compute function and run it to completion to minimize context switches, since a pure function does not block.\footnote{Compute functions do not issue syscalls. \dandelion allocates an isolated memory region for each function before it starts executing, loads inputs in this region (e.g., via a memory-mapped file system), and provides a standard C/C++ library that allows file read/writes without syscalls (see \S\ref{sec:design:interface}).}
 Finally, it enables \dandelion to run a cooperative runtime for network I/O per node (rather than per sandbox) and elastically adjust the CPU core allocation based on the number of compute vs. communication functions in the system.  

We implement several applications on \dandelion, including a distributed log processing application (which interacts with an authentication service and storage servers over HTTP), SQL query processing (which ingests data from S3), and a simple Text2SQL agentic AI workflow (which uses communication functions to prompt an LLM and to query a remote database).
The platform design is not tied to a particular sandbox mechanism. To demonstrate this, we implement \dandelion with four different lightweight isolation backends for compute functions (KVM~\cite{kvm}, Linux processes, CHERI memory capabilities~\cite{cheri}, and rWasm~\cite{rwasm}).
We compare \dandelion's performance and security trade-offs to current FaaS platforms that use Firecracker~\cite{firecracker}, gVisor~\cite{gvisor}, or Wasmtime~\cite{wasmtime} for isolation. \dandelion boots sandboxes in 100s of $\mu$s, or even under 90$\mu$s for CHERI-based sandboxes, which is over an order of magnitude faster than Firecracker snapshot booting. \dandelion makes it practical to cold-start every request, reducing memory overprovisioning by 96\% compared to Knative autoscaling when replaying the Azure Functions trace and reducing execution time variance.
\dandelion improves elasticity for query processing compared to the AWS Athena serverless query service, achieving 40\% lower latency and 67\% lower cost for short-running queries.
The code for \dandelion is available at \url{https://github.com/eth-easl/dandelion}.
\section{Serverless As We Know It Today}\label{sec:background}

FaaS platforms are advertised as elastic cloud services that quickly provision and autoscale server infrastructure for users based on application load. We review their current execution system design, programming model, and discuss how current platforms provide the \textit{illusion} of high elasticity.

\subsection{FaaS Execution Environment} 
FaaS platforms execute and isolate functions by wrapping function code, inputs, outputs, and intermediate execution state inside sandboxes, such as MicroVMs~\cite{firecracker}, hardened containers~\cite{gvisor}, or Wasm modules~\cite{wasm}. Sandboxes are not shared across users or functions~\cite{aws-lambda-exec-environment}. Subsequent invocations of the same function from the same user may reuse a sandbox~\cite{alzayat2023groundhog}. The platform scales the number of sandboxes per function based on invocations~\cite{knative, serverless-in-the-wild}.

Hyperscaler cloud providers use hardware virtualization to sandbox untrusted user code~\cite{nitro, firecracker, azure-security}.
Each function runs in a MicroVM with a guest OS. To breach security, a malicious user function would need to exploit both the guest OS and the hypervisor. The defense-in-depth provided by the VM is critical when user functions can issue a variety of system calls, as the syscall interface is considered a large attack surface prone to vulnerabilities~\cite{linux-kernel-vulnerabilities, kernel-security}. 

Traditional VMs can take tens of seconds to boot~\cite{boxer-elasticity}. Firecracker MicroVMs~\cite{firecracker} reduce boot times to hundreds of milliseconds by removing many general-purpose VMM features that FaaS does not need, such as a BIOS and various device emulators. To further reduce boot time to tens of milliseconds~\cite{catalyzer, reap}, MicroVMs can be restored from snapshots. Unikernel-based VMs~\cite{seuss, unikraft, hand:unikernels, lightvm} reduce boot time to the millisecond range by creating a specialized, single-address-space machine image of the user function with only its necessary kernel functionality.

Some cloud or edge platforms run functions as Wasm modules~\cite{cloudflare-wasm, fastly-wasm, faasm, sledge, spin, hyperlight-wasm}.
Wasm is a binary instruction format designed for a stack-based VM to run code at near-native speed with memory safety guarantees provided by a compiler and runtime~\cite{wasm}.
Wasm sandboxes boot much faster than MicroVMs, which makes it practical to create sandboxes on-demand per request. However, Wasm code does not necessarily run as fast as native code~\cite{lightweight-fault-wasm,understanding-wasm-perf, sledge}. Current Wasm compilers may lack optimizations implemented in more mature compilers like Clang, which can result in more instructions, CPU register pressure, and extra branches~\cite{not-so-fast-wasm}, though this is an active area of research and engineering.
Our experiments in \S\ref{sec:eval:compute} and \S\ref{sec:eval:timeplot} show that Wasmtime~\cite{wasmtime} runs slower than native code for compute-intensive tasks.

\subsection{FaaS Programming Model}
FaaS developers submit code snippets that have a POSIX-like environment and can initiate communication with cloud services.
However, they do not expose the public IP addresses of functions and functions cannot accept network connections~\cite{wawrzoniak2021boxer}. Hence, functions that need to communicate typically exchange data via remote storage~\cite{pocket, locus}.
Most FaaS platforms orchestrate functions as independent, stateless units. Some frameworks allow users to express dependencies and build workflows~\cite{aws-step-functions,aws-airflow,azure-durable-functions}.
However, data dependencies are primarily used as a hint to spin up sandboxes for downstream functions, rather than to optimize data transfers between them. These workflows do not necessarily capture an application's entire dataflow, as an application can still communicate with cloud services within a function~\cite{yu2023following, dataflower, compute-centric-networking}.

\subsection{Why FaaS is not Elastic Today}

Although MicroVM snapshots reduce sandbox boot times down to tens of milliseconds, this latency is still too high to incur for every function invocation. Production traces reveal that many FaaS functions execute for tens of milliseconds or less~\cite{huawei-function-trace, datadog:state21} and would hence experience high overhead.
It is difficult to further reduce boot latency, as we find that at least 8\textit{ms} are spent on loading a minimal snapshot (primarily the guest OS state) by demand paging and on re-establishing the network connection between the host and the guest process (to provide function inputs and retrieve its outputs). Both are necessary in a POSIX-like sandbox environment.

Using Wasm sandboxes is an option to reduce boot times compared to MicroVMs. However, in practice several commercial Wasm-based serverless platforms still apply defense-in-depth strategies for stronger isolation even though some components of the Wasm toolchain have been  formally verified~\cite{veriwasm, Johnson:WaVe:2023, rwasm}. For example, Cloudflare runs multiple instances of their Wasm runtime as separate processes on each machine to separate low-trusted Wasm workers from more trusted ones. They also use Linux namespaces and seccomp for additional sandboxing at the process level~\cite{cloudflare-workers-security}. This limits elasticity, as scaling now involves starting a Wasm runtime (i.e. loading and initialization overheads) as well as setting up a namespace and seccomp filters.

Since quickly initializing a securely isolated POSIX-like environment for cloud user functions is difficult, today's platforms pre-provision sandboxes to reduce the probability of incurring guest OS and networking initialization overheads on the critical path of requests. This gives the illusion of microsecond-scale elasticity at the  expense of consuming extra memory.

Figure~\ref{fig:mem-overhead} compares the committed memory on a server executing a sample of the Azure Functions trace~\cite{serverless-in-the-wild} with Knative's default autoscaling algorithm that provisions extra idle sandboxes (red line) versus the memory usage of active sandboxes over time (blue) line. Running a guest OS inside each function sandbox further adds to the memory footprint~\cite{medes, runD}. Memory usage contributes significantly to the overall cost of operating a FaaS platform, as DRAM dominates server costs~\cite{pond-dram-cost, tmo-dram-cost}. This ultimately leads to higher cost for users.
Since it is usually prohibitively expensive for providers to keep enough sandboxes hot in memory for \textit{all} requests~\cite{firecracker, serverless-in-the-wild, faascache, knative}, some requests incur sandbox creation on the critical path, introducing high performance variability for user applications, as shown in Figure~\ref{fig:hotStartServer}.

\section{\dandelion's Approach}

To achieve high elasticity while eliminating the need to pre-provision idle sandboxes in memory, we propose \dandelion.
Our key insight is that redesigning the traditional POSIX-based programming interface for cloud applications in the cloud-native era enables co-designing a more efficient and elastic cloud platform. We observe that cloud-native applications typically consist of interactions between various cloud services (exposed over REST APIs) and snippets of user-defined compute logic. Hence, we propose a programming model in which applications are expressed as DAGs of \textit{pure compute functions} and \textit{communication functions}.
Pure compute functions are snippets of untrusted user code that consume a declared set of inputs and produce a declared set of outputs.
Communication functions enable interaction with external services.
They are implemented by the \dandelion platform. Users invoke communication functions from user application DAGs, but for security, the implementation of communication functions cannot be modified by users. Currently, \dandelion supports an HTTP communication function that enables interaction with remote services like cloud storage buckets, AI inference services, and data warehouses over REST APIs.  
Writing a \dandelion application involves: 1) expressing the DAG of compute and communication functions using a domain-specific language (DSL) and 2) providing the binaries for user-defined compute functions. \S\ref{sec:design-app} describes the DSL for expressing application DAGs (which also makes function inputs and outputs explicit),  language support for developing user-defined compute functions, and the library of currently supported communication functions.
This programming model is inspired by prior dataflow systems~\cite{spark, beam, naiad, dryad, dryadLINQ, malte-dataflow}. Similar to systems like Spark~\cite{spark}, \dandelion's programming model makes data dependencies and thus data-parallelism and task-parallelism explicit to the platform. In contrast with traditional dataflow graphs, where edges encode internal communication and external communication is only performed at the dataflow sources and sinks, \dandelion function instantiations can either perform local compute or exchange data with other cloud services. Furthermore, we design \dandelion to execute dataflow DAGs with secure isolation in multi-tenant deployments without relying on execution in traditional VMs, which are slow to boot.

Decomposing a cloud application into pure compute and communication functions allows \dandelion to optimize the execution system for high elasticity while maintaining secure isolation guarantees (\S\ref{sec:design:system-arch}).
It enables faster cold starts: pure compute functions can be securely executed in lightweight sandboxes that provide memory isolation (not necessarily VMs), since they do not rely on extra software environments like a guest OS.
The separation of compute and communication also enables \dandelion to use cooperative network I/O and elastically adjust the CPU core allocation based on the number of compute vs. communication functions in the system. The platform can monitor each application's I/O intensity based on its number of compute vs. communication functions. \dandelion leverages this information for CPU core allocation, over-committing CPU cores for communication functions and dedicating CPU cores to compute functions.

\textbf{Target applications.} We design \dandelion for workloads with spiky load patterns (scaling from 0/few to 1/many with long periods of no/low load) or data-dependent parallelism. Many applications can benefit from \dandelion's ability to quickly scale-out to multiple CPU cores when an application computes on many independent inputs and quickly scale-in when the application needs to process data sequentially.
Our goal is to enable fine-grain~\cite{granular-computing}, elastic computing with secure isolation for a variety of cloud-native applications, such as data transformation pipelines (e.g., image compression, video transcoding~\cite{opencv, excamera}), distributed log processing~\cite{minio-log-processing}, CI/CD tasks, elastic query processing (which can include untrusted user-defined function (UDF) execution~\cite{containerized-udf}), and emerging agentic AI workflows~\cite{compound-ai-blog, agentic-ai-databricks}.

\textbf{Non-goals.} However, \dandelion is not designed for applications that require large, frequent state synchronization, like online transaction processing, online gaming services, AI training, or data processing algorithms that rely on fine-grain multithreading and shared memory for performance.
\section{\dandelion Programs}\label{sec:design-app}

\subsection{User Interface}\label{sec:design:interface}

A complete \dandelion program (which we refer to as a ``composition'') is a graph $ G = (V, E) $, where: vertices V are one of (i) user-provided compute-only code, i.e., ``compute functions'', (ii) \dandelion-provided facilities for I/O, i.e., ``communication functions'', or (iii) other \dandelion programs, i.e., ``compositions''.
Directed Edges $E$ = (V1, V2, M) connecting the vertices show that one output set of V1 is an input set of V2. M is a metadata descriptor, specifying which output and input set this edge describes as well as a keyword indicating how data is distributed to instances of V2. The keyword is one of `all', `each', or `key', indicating if all input items in the set should be given to a single instance of V2, if each item should be given to a separate instance, or if items should be given to separate instances grouped by their keys.
Keys are set by the user when formatting output data and are only used for grouping.

\begin{figure}
\centering
\includegraphics[trim={0 0.6cm 0 0},width=0.9\linewidth]{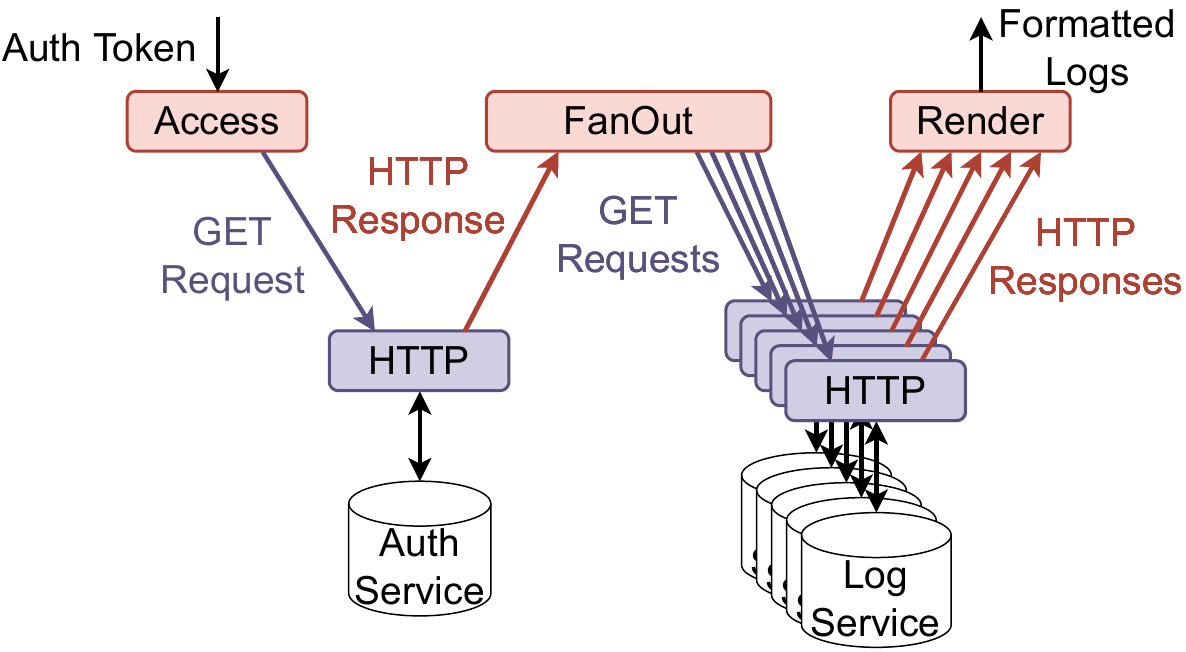}
\caption{Dataflow for log processing example application.}\label{fig:middleware-app-diagram}
\end{figure}

Figure~\ref{fig:middleware-app-diagram} depicts an example \dandelion application that performs  distributed log processing. The composition consists of three user-defined {\color{computation}{compute functions}} and multiple calls to the \dandelion  {\color{communication}{ HTTP communication function}}.
The first compute function, \texttt{\color{computation}{Access}}, takes an input (e.g., an access token) and produces an output (e.g., an http request) that is then passed to the \texttt{\color{communication}{HTTP}} communication function provided by \dandelion. This function carries out the request and forwards the response (e.g., a list of authorized log service HTTP endpoints) to the \texttt{\color{computation}{FanOut}} function. \texttt{\color{computation}{FanOut}} function then takes this list and formats requests to retrieve the log files from each HTTP endpoint, dispatching them to \texttt{\color{communication}{HTTP}} functions in parallel.
The last compute function, \texttt{\color{computation}{Render}}, aggregates the responses and templates them into a single output returned to the user. 
For more complex applications, compositions can include nested compositions, or spawn new compositions dynamically through \dandelion's HTTP interface, e.g., to support dynamic control flow. 

\dandelion provides a composition language to help users express DAGs of compute functions and communication functions in a more developer-friendly syntax.
Listing~\ref{lst:middleware_composition} shows an example of the log processing example application expressed as a composition in \dandelion's DSL. Our DSL is inspired by the domain-specific languages that other dataflow systems like Spark~\cite{spark} and Timely~\cite{naiad} provide for users to specify DAGs.
Our language could be replaced with other DSLs.
Currently, users need to provide the composition and all compute function binaries directly, but ideally applications could be automatically decomposed into pure compute and communication functions, as we discuss in \S\ref{sec:prog:porting}.

\fakepara{Communication Functions.}\label{sec:prog:comm}
Platform-provided communication functions enable user applications to interact with other services. They are designed to support high-level protocols with established user interfaces.
A well-defined user interface enables the platform to safely validate requests to protect against malicious requests. It also frees the user from worrying about the lower-level implementation. Since communication functions are trusted, they are allowed to call system calls to establish and manage new connections. While \dandelion could support protocols such as gRPC or SCP, our current implementation focuses on HTTP, since cloud services generally expose REST APIs~\cite{restful_2013, rest-api}. \dandelion's HTTP function first sanitizes inputs and then makes the specified GET/PUT/POST/DELETE requests, handing the response back to the composition for further processing.

\fakepara{Compute Functions.} \label{sec:prog:compute}
Compute functions can use familiar interfaces for memory allocation, local filesystem operations, and basic utilities like math functions, formatting, etc by linking to \dandelion's custom libc (\dlibc) or libc++ (\dlibcpp). These libraries provide a high-level interface with a userspace in-memory virtual filesystem.
The in-memory filesystem represents function input sets and output sets as folders, with items as files within these folders. This enables a compute function to read inputs and write outputs as standard file operations without invoking system calls. 
Functions requiring system calls (e.g., mmap, mprotect, socket or threading) have stub implementations, returning appropriate error codes.
Our current prototype provides SDKs for developing compute functions in C and C++, as well as a Python interpreter (see \S\ref{sec:prog:sdk}).

\dlibc and \dlibcpp run on top of a lower-level system interface.
This interface uses a special data structure that \dandelion sets up before the function starts to point to descriptors for the input sets and items in the function's memory. Before the function exits, it sets up the structure with the output sets and item descriptors. When a compute function uses the virtual file system to create outputs, \dlibc automatically adds all files inside folders that represent output sets as output items. 

\subsection{Language and Compilation Support}\label{sec:prog:sdk}
\dandelion provides software development kits (SDK) for functions written in C and C++.
These SDKs help users compile their functions against \dlibc /\dlibcpp or the low-level interface.
To support Python functions, we have compiled the CPython interpreter using our C SDK.
The same approach can be taken to support other interpreted languages with interpreters written in C or C++.
To support more compiled langauges, we aim to extend our SDK to be compatible with LLVM, such that any language with an LLVM frontend can be easily compiled for \dandelion.
Additional languages can also be supported by \dandelion's rWasm backend (\S\ref{impl:compute-engine}), as it accepts Wasm code as input, which is a compilation target for many languages~\cite{emscripten}. 

\subsection{Porting Applications to \dandelion}\label{sec:prog:porting}

\dandelion users currently need to express their applications as DAGs of pure compute and communication functions.
Some applications like data analytics naturally fit this model, as query engines represent SQL queries as DAGs of operators. We also find that cloud-native applications like image or log processing pipelines that ingest data from cloud storage fit this programming model well. Based on our initial exploration with a Text2SQL agentic application in \S\ref{sec:eval:target_apps}, \dandelion can also be suitable for some agentic AI workflows, which are an emerging class of applications that involve interactions between AI inference services, vector databases for retrieval, and custom user logic for pre- and post-processing data between these services. 

\begin{lstlisting}[basicstyle=\ttfamily, style=codestyle, caption={Pseudocode for log processing application.},captionpos=b, label={lst:middleware_pseudocode}, float=t]
auth_token = format_request(args[1])
auth_response = http_get(auth_url, auth_token)
authorized_servers = auth_processing(auth_response)
responses = []
for sever in authorized_servers:
    responses.push(http_get(server))
rendered_html = <html template start>
for response in responses:
    if response.successful():
        rendered_html.append(reponse.render())
    else:
        rendered_html.append(response.error())
rendered_html.append(<html template end>)
\end{lstlisting}

However, users may also wish to port existing applications to \dandelion.
To illustrate this process, consider the  pseudocode in Listing~\ref{lst:middleware_pseudocode} for the log processing example application from Figure~\ref{fig:middleware-app}.
To port this application, the developer needs to be split the compute logic into separate functions between each of the two HTTP calls that the application issues. This results in three separate compute functions.
The first compute function, called \texttt{\color{computation}{Access}}, contains the parsing logic for the input token and forms the request (line 1 from Listing~\ref{lst:middleware_pseudocode} ).
The second compute function, called \texttt{\color{computation}{FanOut}}, contains the logic from lines 3-5 in Listing~\ref{lst:middleware_pseudocode} while the third compute function, called \texttt{\color{computation}{Render}}, contains the logic from lines 7-13. 
The HTTP requests issued on lines 2 and 6 in Listing~\ref{lst:middleware_pseudocode} are done by communication functions in \dandelion.
Listing~\ref{lst:middleware_composition} shows the composition that the user would write to express how \dandelion should pass data between compute and communication functions.
The \texttt{\color{blue}{each}} keyword on both of the \texttt{\color{communication}{HTTP}} functions tells \dandelion it can parallelize the requests, while the \texttt{\color{blue}{all}} keywords on the compute functions indicate that all results of the preceding functions are collected and fed to a single instance of the function.

\begin{lstlisting}[basicstyle=\ttfamily, style=codestyle, caption={Log processing application from Listing 1 expressed as a Dandelion composition DAG using our DSL.},captionpos=b, label={lst:middleware_composition}, float=t]
|\color{blue}composition| |\color{computation}RenderLogs|(|\color{teal}AccessToken|) => |\color{brown}HTMLOutput| {
    |\color{computation}Access|(|\color{teal}AccessToken| = |\color{blue}all| AccessToken)
        => (AuthRequest = |\color{brown}HTTPRequest|);
    |\color{communication}HTTP|(|\color{teal}Request| = |\color{blue}each| AuthRequest)
        => (AuthResponse = |\color{brown}Response|);
    |\color{computation}FanOut|(|\color{teal}HTTPResponse| = |\color{blue}all| AuthResponse)        
        => (LogRequests = |\color{brown}HTTPRequests|);
    |\color{communication}HTTP|(|\color{teal}Request| = |\color{blue}each| LogRequests)
        => (LogResponses = |\color{brown}Response|);
    |\color{computation}Render|(|\color{teal}HTTPResponses| = |\color{blue}all| LogResponses)
        => (HTMLOutput = |\color{brown}HTMLOutput|);
}
\end{lstlisting}

For applications that use asynchronous programming~\cite{rust-async} (supported in popular languages like Python, Rust, and C++), we expect that \dandelion could use a compiler to automatically decompose an application into its pure compute and communication components by dividing the application into a graph of continuations~\cite{flanaganEssenceCompilingContinuations1993}.
Continuations are language-level primitives that can save the program state at arbitrary locations (e.g., at communication I/O boundaries) and resume execution at a later time.
We plan to explore this approach as future work, drawing inspiration from VectorVisor~\cite{ginzburgVectorVisorBinaryTranslation}, which uses continuations to port code containing system calls to run on GPUs. 

\subsection{Fault handling}

When a communication function fails, it forwards the appropriate error code to the next downstream function(s).
For example, an HTTP function may return \texttt{404 Not Found} as a response.
\dandelion compositions support conditional control flow, so failures can be handled gracefully.
By default, functions are only executed if each of the input sets contains at least one item.
This allows certain functions to only run if a failure message is present and to skip execution of further downstream functions if the failure cannot be recovered from.
Sets that do not need to contain an item for the function to run can be marked as optional.

\dandelion's failure handling could benefit from additional techniques, such as the log-based approaches in Boki~\cite{jiaBokiStatefulServerless2021}, Beldi~\cite{qiHalfmoonLogOptimalFaultTolerant2023} or Halfmoon~\cite{qiHalfmoonLogOptimalFaultTolerant2023}. These systems can be used to find reentry points in compositions and skip functions that have successfully committed their progress.
\section{\dandelion Execution System}\label{sec:design:system-arch}

We now describe how \dandelion runs a composition and how we optimize the execution system for high elasticity. \autoref{fig:system_arch} shows the \dandelion worker node system architecture. 
A client sends an HTTP request to the frontend.

\fakepara{HTTP frontend.}
The frontend manages client communication, handling requests for composition/function registration and invocation.
It forwards these requests to the dispatcher and serializes and returns the final result to the client.

\fakepara{Dispatcher.}
The dispatcher orchestrates function execution within a \dandelion worker node. 
It keeps track of pending invocations, available compute and memory resources, and maintains a registry of all registered composition DAGs, function binaries, and associated metadata. 
The dispatcher schedules functions by tracking input/output dependencies and determines when a function is ready to run (i.e., when all its inputs are available).

The dispatcher manages memory and compute resources using two abstractions: \textit{memory contexts} and \textit{engines}, respectively. The dispatcher prepares an isolated memory context for each ready function and enqueues the function in a compute or communication engine queue for execution. 
When an engine finishes executing a function, the dispatcher associates its output data to the input of waiting functions. 
When a waiting function becomes ready, the dispatcher ensures that the outputs from prior functions are copied as inputs into the new function's context. 
The dispatcher de-allocates a completed function's memory context when all data-dependent functions have consumed its output.

\begin{figure}
    \centering
    \includegraphics[trim={0 0.5cm 0 0}, width=0.4\textwidth]{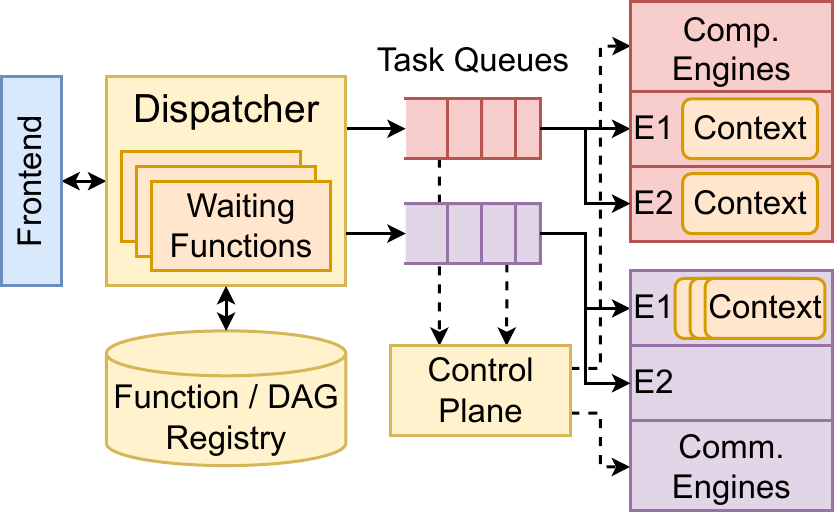}
    \caption{\dandelion worker node system architecture.}
    \label{fig:system_arch}
\end{figure}

\fakepara{Memory contexts.} A memory context is an abstraction for the dispatcher to manage the memory used by a function during execution.
Each context is a bounded, contiguous memory region with methods to read or write at particular offsets and methods to transfer data to other contexts.
Maintaining a simple context abstraction simplifies implementing different isolation mechanisms for compute functions, such as processes or lightweight VMs (\S\ref{sec:impl}).
It also enables specializing methods for transferring data between different types of contexts.
The maximum size of a context is set by the memory requirement specified by the user when registering a function (like in AWS Lambda).
\dandelion reserves this amount of virtual memory for the context and uses demand paging to allocate zeroed pages as needed.

\fakepara{Engines.} 
Engines abstract compute resources that execute functions.
\dandelion implements two types of engines: compute and communication engines. 
Each engine type polls a single, type-specific queue to ensure late binding of tasks.
The dispatcher enqueues tasks (which consist of a prepared memory contexts and metadata) to the appropriate queue type and receives contexts containing the results.

\textit{Compute engines} are responsible for securely executing untrusted user code.
When a compute engine dequeues a task, it sets up the necessary isolation for the memory context and executes the function on the assigned compute resource.
After execution completes, the engine inspects the left over context, updates metadata to indicate the locations of output data, and returns it to the dispatcher. 
\S\ref{sec:impl} describes the implementation details for different isolation mechanisms.
Compute functions do not block, so each compute engine only runs a single task at a time to completion\footnote{Tasks that run for longer than a user-specified timeout (e.g., long or infinite loops) will be preempted to prevent resource hogging.} to minimize interference and context switching.

\textit{Communication engines}, on the other hand, are part of the trusted execution system and therefore do not run in isolated sandboxes. These functions enable interactions (e.g., with HTTP requests) with external services by receiving a context with inputs and carefully validating these inputs to prevent corruption from potentially malicious users (see \S\ref{sec:comm-engines}). They return a context with outputs from the function execution if input validation succeeds or an error if input validation fails.
Unlike compute engines, communication engines use cooperative multi-threading with green threads to maximize efficiency.

\fakepara{Control plane.}
The worker control plane dynamically balances CPU resources between compute and communication engines to maximize application goodput.
It periodically (every 30ms) measures the growth rates of the communication and compute engines' queues. It uses the difference between their growth rates as an error signal for a Proportional-Integral controller~\cite{astrom1995_pid}.
If the control signal is positive, the control plane re-assigns a CPU core from the communication engine type to the compute engine type. If it is negative, it re-assigns a core from the compute engine type to the communication engine type.

\fakepara{Cluster manager.} The cluster manager (not shown in Figure~\ref{fig:system_arch}) orchestrates multiple worker nodes and load balances composition invocations across nodes.
We extended Dirigent~\cite{dirigent} to support \dandelion worker nodes, but other cluster managers could also be used.
\section{Implementation}\label{sec:impl}

We implement \dandelion in $\sim$12K lines of Rust.
We describe how the dispatcher orchestrates execution (\S\ref{sec:dispatcher-impl}), how compute engines provide isolation with support for four different memory isolation mechanisms (\S\ref{impl:compute-engine}), and how we implement communication engines (\S\ref{sec:comm-engines}). 

\subsection{Dispatcher and Engine Coordination}\label{sec:dispatcher-impl}

The dispatcher orchestrates composition invocations using separate green threads~\cite{rust:green-threads}. It queues functions as their inputs become available and coordinates data movement.
To avoid bottlenecks, the dispatcher offloads tasks that are not part of the control flow, such as loading code into contexts, transferring data between contexts, and executing functions.

\textbf{Data passing.} 
To move data between contexts, \dandelion currently copies data. However, this is not fundamental to the programming model.
Different backends could avoid the copy by remapping memory or using copy-on-write when code or data is needed by multiple functions, if the security model allows for it. This is future work.

\subsection{Compute Engines}\label{impl:compute-engine}

Compute engines execute compute functions with one of four memory isolation options. We implement different isolation backends to demonstrate that \dandelion's design is not tied to a particular mechanism. 
Cloud providers can select a mechanism based on its performance and security tradeoffs, which we evaluate in \S\ref{sec:eval} and \S\ref{sec:security}.

\fakepara{KVM-based isolation.}The KVM backend leverages hardware support for virtualization (e.g. Intel VT-x) to isolate compute functions.
Each compute function runs as a lightweight VM without a guest kernel, as we don't need to support syscalls.
To minimize overheads we use an identity mapping between guest virtual and guest physical addresses.
The engine is a minimal hypervisor, which is only responsible for VM launch and exit using Linux's Kernel-based Virtual Machine (KVM) module~\cite{kvm}.
KVM relies on hardware nested paging (e.g. EPT in VT-x) for memory safety.
The memory context is an anonymous memory mapping in the \dandelion process, that is used to back the guest physical address space.
Inspired by Virtines~\cite{virtines}, we circumvent VM setup cost by reusing KVM structures. For this, we reset virtual CPU state and replace the memory backing the guest physical address space, effectively switching the context.

\fakepara{Process-based isolation.}The process-based isolation backend executes each function as a separate process.
\dandelion uses a shared memory region as context.
When a function is ready to execute, the engine that picks up the task spawns a new process.
The process first runs a small amount of trusted code
to map the shared memory into the process.
The new process sends a ready signal to the engine, which uses \texttt{ptrace} to catch and prevent any system calls the compute function may attempt to make. If a syscall is caught, \dandelion will terminate the function and notify the user.
If the function process exits regularly, the engine parses the data left behind in the shared memory to extract the output of the function. 

\fakepara{CHERI-based isolation.} \dandelion can also leverage new CPU hardware security extensions, like CHERI~\cite{cheri}.
With CHERI, load/store instructions take capabilities as inputs rather than integer addresses. 
Instructions succeed if the capability passes permission and bounds checks performed by the CPU hardware, otherwise it faults.
 
\dandelion's CHERI backend can run compute functions as threads within the \dandelion process, since CHERI capabilities enforce isolation within a virtual address space.
We use CHERI's hybrid capability mode~\cite{cheri} to avoid requiring capability compatibility from all user code.
CHERI contexts are contiguous memory regions, pointed to by a code and a data capabilities.
An engine sets these capabilities as default capabilities and ensures it no longer has any other capabilities before jumping into user code.
Special sealed capabilities are used to return from user code, which are guaranteed by hardware to not have been manipulated. 

\fakepara{rWasm-based isolation.} 
\dandelion can also use isolation mechanisms that do not require hardware support.
We show this by implementing a backend using rWasm~\cite{rwasm}, a Wasm-to-safe-Rust compiler.
Developers register compute functions in Wasm code.
\dandelion transpiles the user-defined Wasm code to safe Rust code, then wraps and compiles it to a native shared library.
To execute functions, \dandelion only needs to load the library and call into it using the symbols defined in the wrapper.
Memory isolation is provided by the memory safety guarantees for \texttt{safe} code by the Rust compiler, which is already part of the trusted computing base as \dandelion is implemented in Rust. Since memory isolation is enforced at compile time, this approach does \textit{not} rely on a runtime like Wasmtime for isolation.

\subsection{Communication Engines}\label{sec:comm-engines}

Each communication engine runs a separate kernel thread pinned on a dedicated core, which executes its own asynchronous runtime, using green threads to run multiple requests in parallel. Communication engines have the same interface to the dispatcher as compute engines and similarly poll task queues. Since the communication engines are trusted, they do not need to execute functions in sandboxes, but they do need to sanitize input data, as this is untrusted.
For our HTTP function, we only rely on the first line defined by the protocol to contain the HTTP method and protocol version. \dandelion can check these fields against a fixed set of options and the first part of the URI, which identifies the host to connect to with either a valid IP or a domain name.

\section{Performance Evaluation}\label{sec:eval}

We evaluate \dandelion's sandbox creation (\S\ref{sec:eval:sandbox-creation}) and function execution latency and throughput (\S\ref{sec:eval:compute}) with different isolation backends. We investigate the impact of executing an application as a composition of compute and communication functions in \dandelion versus as single function with Firecracker or Wasmtime (\S\ref{sec:eval:composition}).
We also show that \dandelion's separation of compute and communication functions enables more efficient CPU scheduling (\S\ref{sec:eval:right-sizing}). Finally, we show \dandelion's end-to-end performance benefits when multiplexing real applications with different compute and I/O intensities (\S\ref{sec:eval:timeplot}), more complex workflows for elastic query processing (\S\ref{sec:eval:target_apps}), and its memory savings when running the Azure Functions production trace (\S\ref{sec:eval:azure}). 

\subsection{Methodology}

\textbf{Hardware configurations.} 
We run experiments on several CPU platforms.
Our default machine is a dual-socket Intel Xeon E5-2630v3 with 16 physical cores, 256GiB of DRAM, and a Mellanox ConnectX-3 NIC. We also use this type of machine to run the loader.
For experiments with the CHERI backend, we run \dandelion on an Arm Morello board~\cite{morello} that implements CHERI capabilities on top of the aarch64 instruction set with a 4-core processor, 16GiB of RAM, and a Mellanox ConnectX-3.
We disable frequency scaling for all experiments.
For Azure trace experiments, we use Cloudlab~\cite{cloudlab} d430 nodes, which have dual-socket Intel E5-2630v3 CPUs with a total of 16 physical cores and 64GB of DRAM.

\fakepara{Baselines.}
We compare \dandelion to Firecracker~\cite{firecracker}, the open-source MicroVM hypervisor used by AWS Lambda. 
We use a simple HTTP relay to route requests to MicroVMs and spawn new MicroVMs.
When the relay receives a request for a hot function, it passes it to the first available MicroVM.
When it gets a cold request, it boots a new MicroVM and forwards the request on as soon as that VM is ready. We run Firecracker (which we abbreviate as FC) with and without snapshots~\cite{firecracker:snapshot_restoration}.
The Firecracker baseline represents VM-based isolation approaches. \dandelion's most similar backend for comparison is the KVM backend, as it relies on the same hardware virtualization.

We use the same HTTP relay setup for gVisor~\cite{gvisor}, which uses containers instead of MicroVMs. 
The most comparable isolation backend in \dandelion is the process backend, as it also uses processes with restricted capabilities.

We also run Spin~\cite{spin}, an open-source framework for building and running event-driven applications with Wasmtime~\cite{wasmtime}. We use Spin's recommended default settings, which include enabling pooled allocation with 1000 Wasmtime instances.
For simplicity, we refer to the Spin/Wasmtime setup as Wasmtime or WT.
This baseline represents software fault isolation and is most suitable to compare to \dandelion's rWasm isolation backend.

\fakepara{Software setup.} 
We run Ubuntu 22.04.
On the Morello board, we use a modified kernel with Morello support based on version 6.7.0, with programs running in CHERI hybrid mode.
We use Firecracker 1.5.0, a Firecracker quickstart guest kernel, and a rootfs built on top of Alpine~\cite{alpine} with only necessary packages added.
For gVisor, we use the 20240715 release with an Alpine Linux base image. For best performance, we choose the KVM platform for gVisor’s syscall interception.
For Spin/Wasmtime, we use Spin version 3.0 configured with its recommended defaults and Wasmtime 25.0.3.
We use Dirigent~\cite{dirigent} as the cluster manager for Azure trace experiments.

\begin{table}[t]
\begin{tabular}{|l|c|c|c|c|}
\hline
                    & CHERI & rWasm & process & KVM \\ \hline
Marshal requests    & 12    & 15    & 12    & 30    \\ \hline
Load from disk      & 29    & 147   & 54    & 194   \\ \hline
Transfer input      & 2     & 2     & 6     & 2     \\ \hline
Execute function    & 5     & 20    & 371   & 536   \\ \hline
Get/send output     & 9     & 12    & 9     & 25    \\ \hline
Other               & 32    & 45    & 34    & 102   \\ \hline \hline
Total               & 89    & 241   & 486   & 889   \\ \hline
\end{tabular}
\caption{\dandelion avg. latency breakdown in $\mu$s for each isolation backend running 1x1 matmul on Morello.}
\label{tab:latency-breakdown}
\end{table}

\subsection{Sandbox Creation}\label{sec:eval:sandbox-creation}

Table~\ref{tab:latency-breakdown} shows the unloaded latency breakdown for \dandelion sandbox creation with each isolation backend. The CHERI backend performs best as it executes within a single address space without spawn new threads on the critical path. The main work is loading the executable, parsing it, and moving it to the correct place in memory (\textit{load from disk} in Table~\ref{tab:latency-breakdown}).
The rWasm backend, using software isolation, achieves fast function execution but is mainly limited by slow dynamic loading. 
The process backend spends most of the time creating and preparing new processes (\textit{execute function} in Table~\ref{tab:latency-breakdown}), while KVM backend is slowed by lightweight VM setup, worse on Morello with a CHERI-compatible kernel.
When using the default Linux 5.15 kernel, the total latencies of the rWasm, process, and KVM backends are 109, 539, and 218 microseconds, respectively.

We compare \dandelion's sandbox creation tail latency to other systems as we sweep throughput in Figure~\ref{fig:rps-sweep-morello}, using 1x1 64-bit integer matrix multiplication with 0\% hot invocations. Creating a sandbox in Firecracker (FC) involves booting a fresh MicroVM, which takes over 150ms. FC can reduce startup latency by loading pre-initialized MicroVM state from a snapshot on disk instead of initializing the guest OS and application on the critical path. However, snapshot restoration still involves creating a MicroVM and restoring its state, limiting it to 120 RPS.
gVisor~\cite{gvisor} uses a hardened container instead of a MicroVM. 
As gVisor performed worse than FC with snapshots, we omit it in remaining plots.
Initializing a Spin/Wasmtime (WT) instance is lightweight due to multiple optimizations, including pooling memory allocation and pre-instantiating Wasm components, which allows WT to achieve 7000 RPS peak throughput.
Although not shown in the figure, we also experimented with Microsoft's recent Hyperlight Wasm~\cite{hyperlight-wasm} system, using their default sandbox configurations.
We measure 9.1 ms average unloaded cold start latency with Hyperlight Wasm, which includes launching a ProtoWasmSandbox (2.8ms), loading the Wasmtime runtime (4.2ms) and loading the Wasm module (2.1ms).
We also considered Unikraft~\cite{unikraft}, but did not run separate experiments. Their paper reports a 3.1ms VM boot time to user space \textit{main} function using Firecracker as VMM.
This excludes the time to receive a request and send a response, which would be inclued in our experiment, making its performance very similar to FC with snapshots.

\begin{figure}[t]
    \centering
    \includegraphics[trim={0 1cm 0 0},width=0.95\columnwidth]{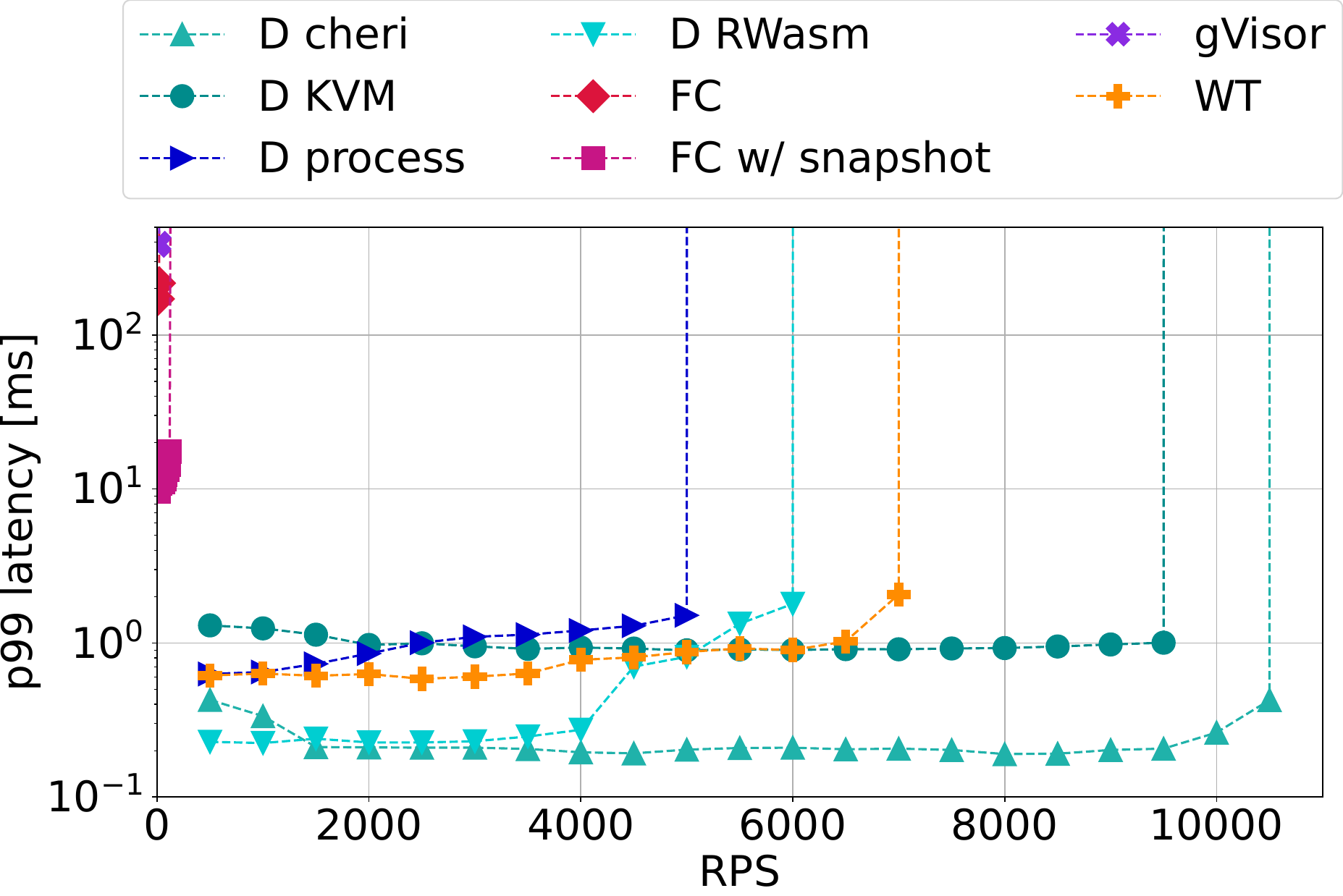}
    \caption{Tail latency vs. throughput for 1x1 matmul on the Morello server, with 0\% hot requests.}
    \label{fig:rps-sweep-morello}
\end{figure}

\subsection{Compute Function Performance}\label{sec:eval:compute}

Figure~\ref{fig:rps-sweep-97hot} shows a compute microbenchmark (128x128 64-bit integer matrix multiplication) on the default server setup.
We conservatively use a 97\% hot request ratio for Firecracker experiments, based on Shahrad et al.~\cite{serverless-in-the-wild} observing that 3.5\% of applications in the Azure Functions trace experience \textit{only} cold starts, even if sandboxes are never torn down.
We plot median latency with 5th and 95th percentile error bars to show the impact of cold starts on overall performance.
\dandelion and Wasmtime create a new sandbox per request. 
For \dandelion, we load the function binary from disk instead of in-memory cache for 3\% of requests.
\dandelion shows low, stable latency, peaking at 4800 RPS with the KVM backend, but its rWasm backend suffers from slower matrix multiplication code due to transpilation. 

\begin{figure}[t]
    \centering
    \includegraphics[trim={0 1cm 0 0},width=0.9\columnwidth]{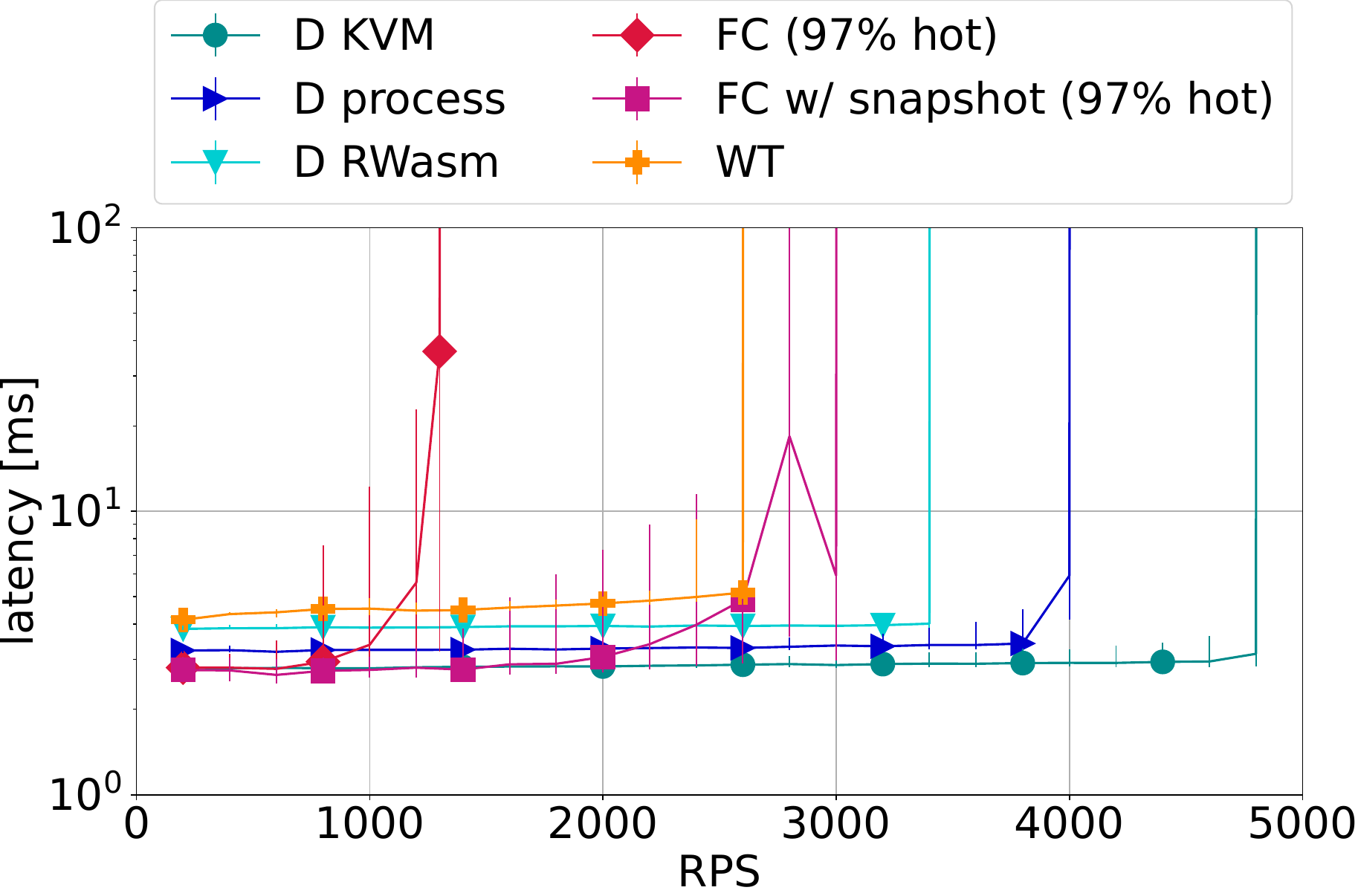}
    \caption{Compute function (128x128 matmul) on 16-core server. Latency is median, with 5\textsuperscript{th}/ 95\textsuperscript{th} percentile error bars.  \vspace{-1.6em}}
    \label{fig:rps-sweep-97hot}
\end{figure}

Firecracker shows similarly low unloaded latency by committing extra memory for hot MicroVMs, ensuring that most requests do not require booting a new sandbox. However, its latency gradually increases and saturates at 3000 RPS (with snapshots), becoming increasingly unstable after 2800 RPS due to CPU contention between active MicroVMs and MicroVM creations.
Wasmtime, despite its fast sandbox creation, has higher unloaded latency and saturates at only 2600 RPS due to the compiler toolchain's less efficient code generation and optimization.
To support 128x128 matrix multiplication in Hyperlight Wasm, we increase default configuration guest\_input\_buffer, guest\_output\_buffer, and guest\_stack\_size to 1 MB each, and guest\_heap\_size to 2 MB, which is the minimum required to avoid out of memory errors. Hyperlight Wasm has a high unloaded average latency of 27.5ms, including sandbox creation (2.6ms), runtime loading (12.1ms), module loading (4.7ms), and function execution (8.1ms). 

\subsection{Composition Performance Overhead}\label{sec:eval:composition}

While \S\ref{sec:eval:sandbox-creation} showed performance of individual sandbox creation, decomposing applications into compute and communication functions can increase the number of sandboxes that need to be created.
To evaluate this tradeoff, we design a microbenchmark that fetches a 64KiB array and computes sum, min and max over a sample of the elements; we call this sequence a phase. We sweep the number of phases in the microbenchmark from 2 to 16.

We compare \dandelion KVM with Wasmtime, Firecracker hot (VM already running) and Firecracker with cold start from snapshots.
For \dandelion, we consider loading the compute function code from disk (uncached) and copying from memory (cached) to see how keeping code ready in RAM impacts chains of functions.
All systems show linear growth in latency as we increase the number of compute-communication phases.
Firecracker hot scales best, but \dandelion KVM uncached is only 1.3ms (17\%) slower than Firecracker hot at 8 phases and 4ms slower at 16 phases. 
Notably, \dandelion KVM cached vs. uncached differs by just 0.5ms at 16 phases, indicating disk loading is competitive even for long chains. 
Wasmtime has the same slope as \dandelion KVM, hinting at potentially worse application scaling on Wasmtime, which is balanced out by the repeated sandbox creations in \dandelion in this experiment.
Firecracker cold start from snapshots has higher base latency but same latency slope as Firecracker hot.
Nonetheless, \dandelion KVM uncached is 4.6x faster at 16 phases. 
Overall, our results show that \dandelion still provides great performance even for applications requiring numerous sandbox creations.

\begin{figure}[t]
	\centering
    \includegraphics[width=0.44\textwidth]{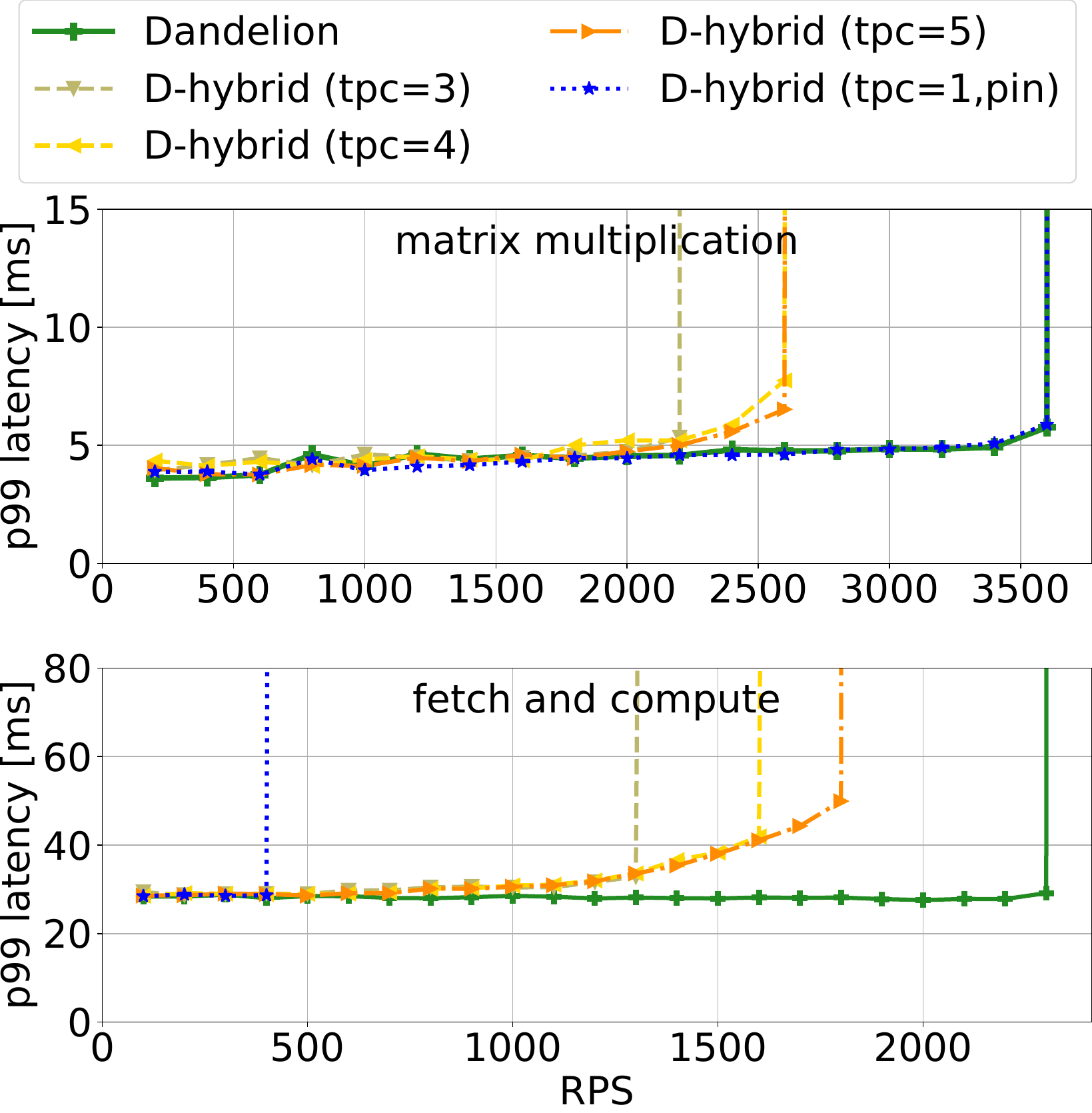}
        \vspace{-0.4cm}
	\caption{Separating compute and communication function execution in \dandelion vs. unifying in \hybrid for two types of applications (compute-intensive and I/O intensive).}
	\label{fig:dandelion-cpu-efficiency}
\end{figure}

\subsection{Throughput Benefits of Compute-Comm Split}\label{sec:eval:right-sizing}

In this section, we highlight an additional benefit \dandelion gains from splitting compute and communication: more efficient core multiplexing.
In traditional FaaS, each sandbox runs its own networking, and the OS scheduler multiplexes sandboxes across CPU cores, incurring costly context switches.
Determining the optimal number of sandboxes to execute concurrently per node to maximize application throughput is hard. 
Compute-intensive sandboxes benefit from low CPU oversubscription, while I/O intensive ones need high oversubscription to fully utilize the CPU resources. Moreover, sandboxes can switch between being I/O intensive and compute intensive at any time.

\dandelion's compute-communication split enables cooperative networking, run-to-completion for compute functions, and efficient reallocation of CPU cores between compute and communication functions. 
To measure the impact of \dandelion's programming model, while keeping the rest of the system the same, we implement \dandelion -hybrid (\hybrid). It uses the same system architecture and isolation backends as \dandelion, but supports running a composition as a single ``hybrid'' function, allowing opening sockets for communication.

In Figure~\ref{fig:dandelion-cpu-efficiency}, we run the compute-intensive (matrix multiplication) and I/O-intensive (fetch and compute) microbenchmarks used in \S\ref{sec:eval:composition}, varying 1 to 8 threads per CPU core (tpc), with and without core pinning for D-hybrid.
We present the highest-performing configurations in either microbenchmark (3-5 tpc without pinning and 1 tpc with pinning).
Depending on the compute and I/O intensity of the applications, D-hybrid requires fundamentally different levels of concurrency to maximize throughput (i.e. 1 tpc pinned for matrix multiplication and 5 tpc unpinned for fetch and compute).
Designing a controller to adjust concurrency based on system load metrics like CPU utilization is non-trivial: matrix multiplication achieves peak throughput at 90\% CPU utilization with tpc 1 with pinning, while fetch and compute achieves peak throughput at 70\% CPU utilization with tpc 5 without pinning.

In contrast, \dandelion's control plane achieves the highest
throughput for both workloads.
\dandelion delivers even higher throughput and consistently lower tail latency for the I/O-intensive application by running compute tasks to completion on dedicated cores and aggregating cooperating communication tasks on the remaining cores, minimizing context switches.
In the next section, we further show how \dandelion multiplexes requests from multiple applications with different compute and I/O intensities.

\begin{figure}[t]
    \centering
  \includegraphics[trim={0 0cm 0 0},width=0.45\textwidth]{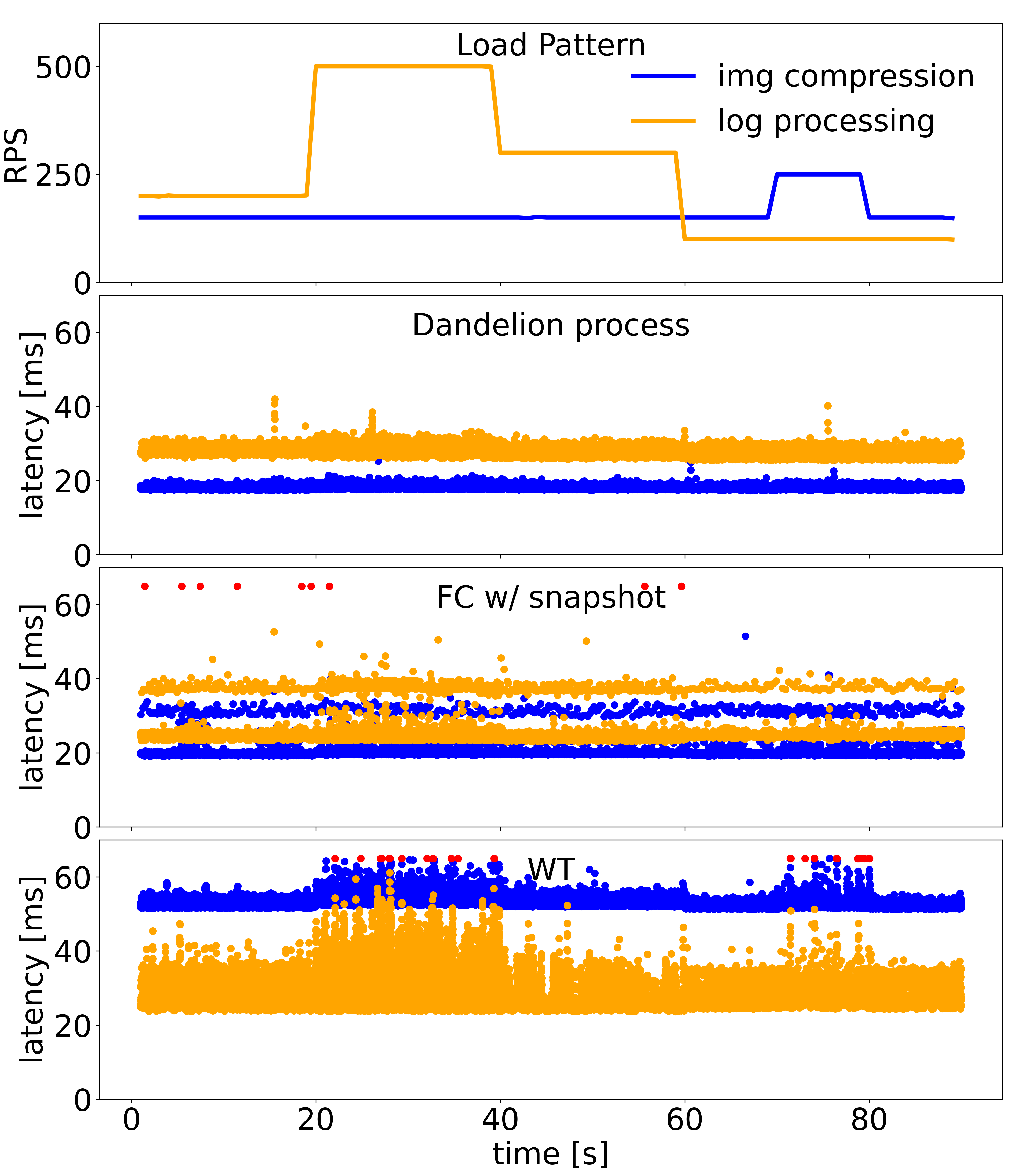}
    \caption{Multiplexing compute vs. I/O-intensive apps, with changing RPS over time. \dandelion dynamically adjusts compute/comm engines to maintain low latency. Red dots show requests where latency exceeds 65ms.
    \vspace{-0.8em}}
    \label{fig:middleware-app}
\end{figure}

\subsection{Multiplexing Applications}\label{sec:eval:timeplot}
While previous experiments showed that \dandelion outperforms Firecracker and Wasmtime with compute and composition microbenchmarks, we now evaluate its behavior in a more realistic setting where the system needs to manage mixed workloads with bursty traffic.

In Figure~\ref{fig:middleware-app}, we run the distributed log processing application from Figure~\ref{fig:middleware-app-diagram} (I/O-intensive) and an image compression application that transforms an 18kB QOI~\cite{qoi-compression} image to PNG (compute-intensive) on \dandelion, Firecracker (with snapshot), and Wasmtime under a bursty load pattern. Image compression is as an example of image processing and log processing is an example of HTML templating, which are both common task from FaaS benchmarks~\cite{copikSeBSServerlessBenchmark2021, kimFunctionBenchSuiteWorkloads2019, vSwarm}.
Consistent with previous experiments, we use a 97\% hot request ratio for Firecracker.

Firecracker has a bi-modal latency distribution for both applications. This is because 97\% of requests are served by hot VMs, which benefit from pre-warmed state, while the remaining 3\% incur cold-start (snapshot restoration) overhead.
By overprovisioning memory to maximize warm starts, Firecracker achieves a relatively low average request latency for both compression (20.4ms) and log processing (25.6ms), but has high variance at 389\% and 1495\% respectively due to the cold starts.
Spin/Wasmtime, on the other hand, struggles to handle the mixed workload effectively, increasing the latency of both applications.
As discussed in \S\ref{sec:eval:compute}, Wasmtime suffers from less efficient code generation and optimization. 
Spin’s reliance on the Rust Tokio async runtime also exacerbates interference between applications~\cite{tokio-runtime, tokio-scheduling-blog}.
Spin schedules functions cooperatively, meaning a thread can be occupied by a compute intensive function for an arbitrary amount of time.
We observe that image compression tasks hog resources, causing log processing tasks to queue behind them, increasing tail latency.
While this keeps the variance of the compression application relatively low at 6.1\%, this is not a good tradeoff as the average latency is 53.3ms. The log processing application has a 79.2\% variance with an average latency of 28.9ms.

\dandelion maintains stable latency for both applications, even during load bursts, due to fast cold starts and the controller's dynamic re-allocation of CPU cores, e.g., it scales from a single I/O core at the beginning of the experiment to 4 I/O cores to accommodate I/O-intensive log processing during peak load. 
For the image compression application, \dandelion achieves an average latency of 18.2ms and p99 latency of 19.6ms, which is 38.6\% lower than Firecracker and 67.1\% lower than Wasmtime.
For the log processing application, \dandelion delivers an average latency of 27.9ms and p99 latency of 30.6ms, which is 19.5\% lower than Firecracker and 33.5\% lower than Wasmtime.
\dandelion reduces performance variability, achieving the lowest relative variance of 1.3\% for image compression and 2.9\% for log processing.
Overall, \dandelion efficiently multiplexes requests with different compute and I/O intensities and maintains low and stable latency despite creating a sandbox on the critical path for each request.



\subsection{Elastic Data Processing with \dandelion}
\label{sec:eval:target_apps}

We now explore \dandelion's suitability for a broader range of target applications with more complex workflows.

\begin{figure}[t]
    \centering
    \includegraphics[trim={0 1.25cm 0 0},width=0.45\textwidth]{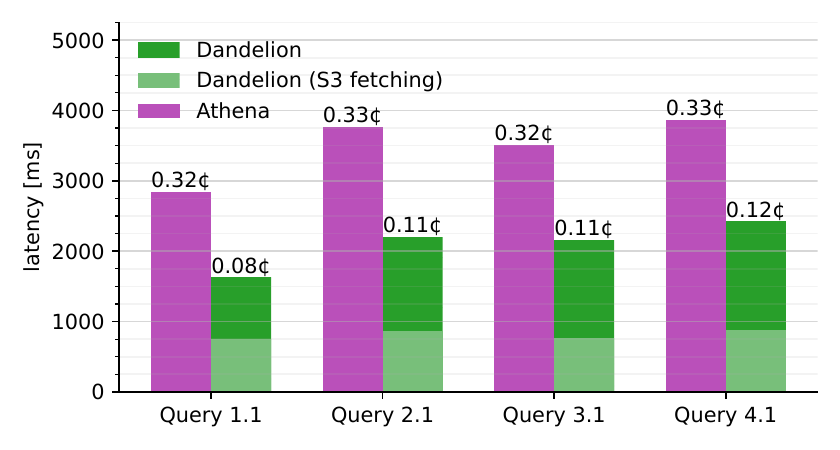}
    \vspace{1em}
    \caption{SSB query latency and cost in US cents compared to Athena. The queries ingest $\sim$700MB input data from S3. 
    } 
    \label{fig:ssb_query_latency}
\end{figure}

\fakepara{Serverless query processing.} Query-as-a-Service (QaaS) offerings like AWS Athena~\cite{athena} enable SQL-based data analysis on cloud storage, with providers managing infrastructure and billing per byte read.
We show that \dandelion improves QaaS elasticity.
Figure~\ref{fig:ssb_query_latency} shows query execution latency and cost with Athena and \dandelion for Star Schema Benchmark~\cite{ssb-benchmark} queries (which are based on the industry standard TPC-H~\cite{tpch} benchmark) using 700MB of input data. The queries include filter, projection, join, order by, and aggregation operators, which we implement in \dandelion by porting the Apache Arrow Acero~\cite{acero} library operators and linking against \texttt{hlibc++}. We run \dandelion on an EC2 m7a.8xlarge instance in the same region as the S3 bucket and compute cost based on the execution time and EC2 VM price. Athena charges per bytes read and we exclude the queuing delay in the plot.
\dandelion quickly boots sandboxes and spreads query execution across all 32 CPU cores, achieving 40\% lower latency and 67\% lower cost than Athena. We expect \dandelion's secure isolation of query operators to be particularly useful for queries that embed user-defined functions (UDFs), as these contain arbitrary and untrusted user code that needs to be sandboxed in multi-tenant environments~\cite{containerized-udf}. For example, AWS Athena executes UDFs in AWS Lambda.
While this experiment does not compare the two systems running on the same hardware or with the same operator implementations, we include it to compare the latency and cost of a commercially available service with what we can achieve using \dandelion as the underlying QaaS system on EC2 VMs. 
With larger input data sizes (we tested up to 7GB), matching Athena's latency requires scaling query execution across multiple \dandelion nodes, but we continue to see lower query execution cost compared to Athena.

\fakepara{Agentic AI query processing workflow.} We explore \dandelion's suitability for emerging agentic AI applications, which consist of multiple interacting components, including model inference services, retrievers, and external tools and often require pre- and post-processing compute logic between these calls~\cite{agentic-ai-databricks, efficient-compound-ai-hotos}. 
We target Text2SQL workflows, a popular subset of agentic AI application, converting natural language queries to SQL.
A Masters student with no prior experience with \dandelion ported a Text2SQL workflow from the TAG benchmark suite~\cite{biswal2024text2sqlenoughunifyingai} in a few hours.
The workflow processes a sample prompt in $\sim2$s with five steps:
1. parsing the input prompt (221 ms), 2. requesting an LLM with the prompt via HTTP (1238 ms), 3. extracting the SQL query from the LLM’s response (207 ms), 4. issuing the SQL query via HTTP to a SQLite database (136 ms), and 5. formatting the database response (213 ms). 
We implement the parsing, SQL extraction, and formatting as \dandelion Python compute functions, while the LLM and database queries are performed by \dandelion communication functions. 
We use Gemma-3-4b-it running on a single H100 NVL as our LLM inference service exposed over a REST API. 
The LLM inference step is the pipeline's bottleneck, accounting for 61\% of the total end-to-end latency.

\subsection{Azure Functions Trace Evaluation}\label{sec:eval:azure}

\begin{figure}[t]
    \centering
    \includegraphics[trim={0 1cm 0 0},width=0.5\textwidth]{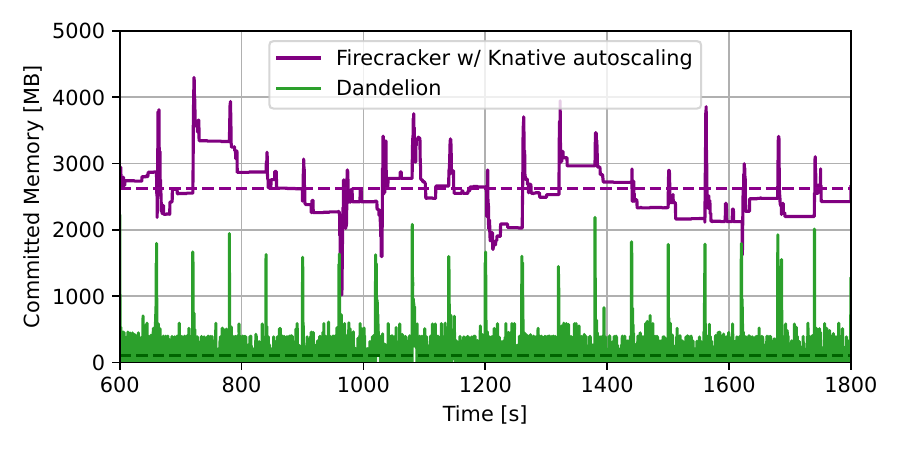}
    \caption{Memory committed over time for Azure trace experiment. Dotted lines show the average. 
    } 
    \label{fig:azure-trace}
\end{figure}

Finally, we compare the performance and memory efficiency of \dandelion (with process isolation backend) and Firecracker, while running the Azure Functions trace sample in the CloudLab setup. We sample 100 functions from day 6, hour 8 of the trace using the InVitro~\cite{ustiugov:in_vitro} sampler.
We use the autoscaling policy in Knative~\cite{knative}, a popular open-source FaaS orchestrator, to control the number of Firecracker MicroVMs over time based on application load.
We find that Knative scales MicroVMs for this trace such that on average 3.3\% of invocations are cold. This validates our 97\% hot requests setting in prior experiments.

Figure~\ref{fig:azure-trace} shows the memory committed by Firecracker with Knative autoscaling compared to \dandelion over a 20-minute long trace. To minimize cold starts, Firecracker maintains MicroVMs long after a request has finished executing, which requires committing memory for long periods. In contrast, \dandelion commits and consumes memory only while requests are actively running since a new context is created for each request. This enables \dandelion to commit only 4\% of the memory Firecracker commits on average (109 MB vs. 2619 MB), achieving significant cost savings and comparable or even lower end-to-end response time for functions. 
\dandelion also reduces the p99 end-to-end latency by 46\% 
compared to Firecracker in this experiment. 

\section{Security Analysis}\label{sec:security}

To compare the security implications of \dandelion and systems like Firecracker, gVisor, and Wasmtime, we analyze their attack surface and trusted computing base (TCB).

\fakepara{Threat model.} We assume users trust the cloud platform ,but not other users. The cloud provider does not trust users. Users do not have physical access to machines, but can exploit vulnerabilities in cloud software~\cite{linux-kernel-vulnerabilities, hypsec}
to breach the confidentiality and integrity of other users' applications.
Denial of service and side-channel attacks are out of scope; we assume the provider deploys necessary defenses~\cite{microsoft-prevent-sidechannels}.

\fakepara{Attack surface.} We analyze the vulnerability of the interface that untrusted functions can directly access. Most FaaS platforms provide user functions a POSIX-like environment with various system calls available. To mitigate this large attack surface~\cite{firecracker-container-escape}, Firecracker and gVisor rely on defense-in-depth strategies, running system calls for different functions with separate OS kernels.
Wasmtime offers a more controlled attack surface via WASI~\cite{wasi}.
\dandelion goes further in this direction: users provide only pure compute functions and system calls are blocked during their execution (\S\ref{sec:prog:compute}).
Although a malicious compute function could produce output data that attempts to hijack control flow when parsed by the dispatcher,
\dandelion's function output parser is merely 100 lines of Rust, making it feasible to verify its memory safety and correctness. 
Users may attempt to exploit communication functions by passing malicious inputs. \dandelion's communication engines guard against such attacks by checking that each input is a valid HTTP protocol request.

\fakepara{Trusted computing base.}
\dandelion, Firecracker and Spin have several TCB components in common: the Linux kernel, C standard library, and Rust compiler. 
While the Firecracker and Spin code base has expanded to $\sim$68k and $\sim$65k lines of Rust code respectively, \dandelion consists of $\sim$12k lines of Rust code including tests, with only 2k lines directly relevant to isolation and interaction with user code.
gVisor has $\sim$38k lines of Go, excluding third-party packages.
\section{Related Work}\label{sec:relatedwork}

\textbf{Serverless programming models.} \dandelion's design is closely related to other dataflow paradigms for serverless. Computation-centric networking~\cite{compute-centric-networking} involves a separation between I/O and compute, with delineated nondeterminism. Lambdata~\cite{lambdata} and Dataflower~\cite{dataflower} leverage application dataflow information to reduce function triggering latency and avoid round-trips to remote storage for data exchange.
Pheromone~\cite{yu2023following} exposes a data bucket abstraction (for intermediate function outputs) and trigger primitives, which allows developers to specify when and how to pass intermediate data to other functions and invoke them. 
Other FaaS systems introduce data dependency hints to optimize function placement across nodes~\cite{abdi2023palette, mahgoub:sonic, faasflow}. 
Nu~\cite{nu-nsdi} provides a new process abstraction that decomposes an application into fine-grained units of state and compute, which can be quickly migrated to improve resource utilization. However, Nu only provides process-level isolation, which is not sufficient for multi-tenant cloud environment.

SigmaOS~\cite{sigmaos} forgoes POSIX compatibility and designs a cloud-centric communication API to improve container performance for serverless and microservice tasks. However, SigmaOS still allows user code to invoke 67 host system calls, which has different security implications than Firecracker, gVisor, and \dandelion, which restrict user code from making direct system calls to the host kernel.

\fakepara{Lightweight isolation mechanisms.}
Complementary to \dandelion's four isolation backends, Wedge~\cite{wedge-nsdi08} introduces sthreads for intra-process privilege separation and memory isolation, Shreds~\cite{shreds-security16} isolates code and data compartments using Arm memory domains, and lwCs~\cite{lwC-osdi16} are independent units of protection, privilege, and execution state within a process.
Virtines~\cite{virtines} is a programmer-guided abstraction allowing individual functions within an application to run in lightweight, virtualized execution environments. X-containers~\cite{x-containers} is an exokernel-inspired container architecture designed to isolate single-concerned cloud-native applications without nested hardware virtualization.
\mbox{GraalOS~\cite{graalos-serveless}} and V8~\cite{v8} isolate code with language runtimes.

\fakepara{Data processing systems.}
DAG-based systems for data processing such as Spark\cite{spark} are already popular.
These systems usually focus on single users or groups of mutually trusting users. 
\dandelion innovates on the DAG model for processing, as it leverages it to achieve secure isolation and high elasticity in multi-tenant environments.

\section{Conclusion}\label{sec:conclusion}

The cloud has evolved in the past decade to offer a variety of serverless compute, storage, and AI/analytics services, but the performance and efficiency of these services is limited by the legacy software that they retrofit. Today's infrastructure is still based on the older, more traditional model of cloud computing, in which users rent long-running VMs, each with a guest OS exposing a POSIX-like interface. Initializing this system stack is slow. Furthermore, having applications interact with the platform on the level of files and sockets does not provide the platform with useful information about application characteristics, like compute vs. I/O intensity, to optimize resource allocation and scheduling.

\dandelion co-designs a declarative cloud-native programming model and execution system for truly elastic computing while maintaining secure isolation. 
Developers specify their applications as DAGs of pure compute functions and communication functions.
This enables \dandelion to isolate pure compute functions with lightweight sandboxes that boot in 100s of microseconds, enabling true elasticity (i.e., fast cold start for every request). \dandelion improves elastic query processing latency by 40\% compared to AWS Athena and reduces committed memory by 96\% on average compared to Knative autoscaling for the Azure Functions trace.

\section{Acknowledgements}

We would like to thank Gustavo Alonso, Timothy Roscoe, Xiaozhe Yao, and Maximilian Böther for their helpful insights during discussions. We would also like to acknowledge
Roberto Starc, Lorenzo Rosa, Georgijs Vilums, and Josef Schönberger, and Alessio Russo for their contributions to the project. Thank you to Eric Van Hensbergen for his inputs in the early stages of the project and his support with the Arm Morello setup.
We also thank our anonymous reviewers from ASPLOS'25 and OSDI'25 for their feedback and suggestions. Tom Kuchler, Pinghe Li, and Yazhuo Zhang are supported by the Swiss National Science Foundation (project TMSGI2\_218019).
\bibliographystyle{plain}
\bibliography{references}

\end{document}